\documentclass[amsmath,amssymb,aps,10pt,prd,letterpaper,nofootinbib,balancelastpage,notitlepage,superscriptaddress,twocolumn,floatfix,preprintnumbers]{revtex4-2}
\usepackage{graphicx}	
\usepackage{ragged2e}	
\usepackage{bm}		    
\usepackage[sort&compress]{natbib}	 	
	\setcitestyle{square,numbers,comma}	
\usepackage[colorlinks=true,urlcolor=blue,linkcolor=black,citecolor=red]{hyperref}	
\usepackage{slashed}

\renewcommand{\eqref}[2][]{Eq{#1}.~(\ref{eq:#2})}		
\newcommand{\orcid}[1]{\href{https://orcid.org/#1}{\,\includegraphics[width=8px]{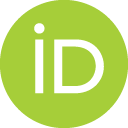}}}

\usepackage{color}
   \definecolor{k}{RGB}{0,0,0}
   \definecolor{r}{RGB}{255,0,0}
   \definecolor{R}{RGB}{255,0,0}
   \definecolor{o}{RGB}{255,94,1}
   \definecolor{O}{RGB}{255,165,0}
   \definecolor{p}{RGB}{148,55,255}
   \definecolor{P}{RGB}{148,55,255}
   \definecolor{g}{RGB}{0,150,50}
   \definecolor{G}{RGB}{0,150,50}
   \definecolor{b}{RGB}{0,0,255}
   \definecolor{B}{RGB}{0,0,255}

\begin{document}

\title{Lunar radio emission from dark photon dark matter}

\author{Ariel Arza\orcid{0000-0002-2254-7408}}
\email{ariel.arza@gmail.com}
\affiliation{Department of Physics and Institute of Theoretical Physics, Nanjing Normal University, Nanjing 210023, China}
\affiliation{Nanjing Key Laboratory of Particle Physics and Astrophysics, Nanjing, 210023, China}

\begin{abstract}
The dielectric surface of the Moon provides an efficient interface for the conversion of dark matter dark photons into radio waves. This work explores a novel detection strategy that aims to capture this radio emission using standard radio telescope antennas. Detailed sensitivity projections are presented that demonstrate the power of this technique. It is found that both current and future radio observatories, including large arrays such as the Atacama Large Millimeter/submillimeter Array (ALMA) and the Square Kilometer Array (SKA), as well as large single dishes like the Five-hundred-meter Aperture Spherical Telescope (FAST) and the Green Bank Telescope (GBT), can explore a substantial and previously inaccessible parameter space for these dark matter candidates.
\end{abstract}

\maketitle

{\it \textbf {Introduction}.} Ultralight bosonic fields are compelling dark matter candidates \cite{Kimball:2023vxk}. Produced non-thermally in the early universe \cite{Arias:2012az}, these fields can attain high occupation numbers \cite{Sikivie:2009qn} and consequently behave, in part, as coherent classical waves \cite{Sin:1992bg}. A distinctive feature of such fields is their ability to convert into electromagnetic radiation. This phenomenon is well established for axions and axion-like particles in the presence of external magnetic fields \cite{Sikivie:1983ip}. Similar conversion processes also arise for other well-motivated ultralight dark matter candidates, including dark photons \cite{Okun:1982xi,Holdom:1985ag}, which can likewise be described as coherent classical fields \cite{Caputo:2021eaa,Jaeckel:2012mjv,Jaeckel:2010ni}. The cosmological abundance of dark photon dark matter depends on its production mechanism, with possibilities including inflationary fluctuations \cite{Graham:2015rva}, parametric resonance from a dark Higgs \cite{Dror:2018pdh}, gravitational production \cite{Sato:2022jya}, and tachyonic instabilities sourced by inflaton or axion-like fields \cite{Bastero-Gil:2018uel,Co:2018lka}.

Ultralight dark photons can efficiently convert into electromagnetic radiation at conducting and dielectric boundaries \cite{Jaeckel:2013eha}. In conducting materials, this principle underlies searches using dish antennas \cite{BREAD:2023xhc,Ramanathan:2022egk,BREAD:2021tpx,Brun:2019kak,Suzuki:2015sza,Horns:2012jf} and spherical mirrors \cite{FUNKExperiment:2020ofv}. Conversion can also occur in dielectric materials, such as at the surfaces of stacked dielectric plates, forming so-called dielectric haloscopes \cite{Caldwell:2016dcw,Baryakhtar:2018doz,Koppell:2025dmt}. These experiments have enabled dark photon dark matter searches and have placed new constraints on the corresponding parameter space \cite{MADMAX:2024jnp,Manenti:2021whp,Wei:2025kgb,Li:2026iso}.

A natural extension of this conversion principle is provided by the Moon, whose surface can act as a large dielectric converter for dark photon dark matter. Dark photon conversion at the lunar surface produces electromagnetic radiation that is emitted predominantly away from the surface, allowing part of the signal to propagate toward Earth and be detected with radio telescopes. This approach effectively repurposes existing astronomical instruments as dark matter detectors. Radio observations have previously been considered as probes of ultralight dark matter. For axions, proposed and existing searches include axion decays into photons in Local Group dwarf galaxies \cite{Blout:2000uc} and resonant axion-photon conversion in the magnetospheres of neutron stars \cite{Foster:2020pgt}. Dark photon dark matter has similarly been investigated through resonant conversion to photons in the solar corona \cite{an:2023wij,An:2024wmc}, radio emission from radio telescope surfaces \cite{An:2022hhb}, and proposed observations of the Galactic Center \cite{Arza:2024iuv,Hardy:2022ufh}.

The Moon provides a distinctive conversion target because of its enormous naturally occurring surface area. Whereas laboratory haloscopes are limited by the physical dimensions of their conversion surfaces, the lunar surface provides a converting area of order ten million square kilometers. The corresponding advantage is partly offset by the Earth--Moon distance and by the finite collecting area of radio telescopes. Consequently, the sensitivity is governed by the combined effects of the conversion area, telescope collecting area, and source distance, yielding a figure of merit that can be comparable to that of laboratory haloscopes. The lunar surface therefore provides a complementary conversion target that requires no dedicated detector hardware beyond an existing radio telescope. In addition, radio telescopes can record broad frequency ranges in a single observation, allowing a narrow spectral-line signal to be searched for across a broad range of dark photon masses without the sequential tuning required by conventional haloscopes.

The electromagnetic flux density radiated from the Moon is computed while accounting for the effects of lunar surface roughness over length scales ranging from fractions of a millimeter to several meters, as characterized by measurements from lunar missions. Based on this calculation, sensitivity projections are derived for large single-dish radio telescopes, including the Five-hundred-meter Aperture Spherical Telescope (FAST) and the Green Bank Telescope (GBT), as well as for large antenna arrays, including the Atacama Large Millimeter/submillimeter Array (ALMA) and the future Square Kilometre Array (SKA). This approach enables a search for dark photon dark matter over a broad frequency range from $50\,\mathrm{MHz}$ to $950\,\mathrm{GHz}$, corresponding to dark photon masses of order $(\mathrm{few})\times10^{-7}$--$(\mathrm{few})\times10^{-3}\,\mathrm{eV}$.
\newpage

{\it \textbf {Signal from the lunar surface}.} At the interface between two regions with different dielectric constants, dark photon dark matter induces a time-dependent electric field that is discontinuous across the interface. The requirement that the parallel components of the electric field satisfy the appropriate continuity conditions leads to the emission of an electromagnetic wave from the interface. This mechanism applies to the lunar surface, which consists of dielectric material with a dielectric constant $\epsilon$ ranging from approximately 1.5 to 5, depending on the lunar region. In the absence of an atmosphere, the medium above the lunar surface is vacuum, with dielectric constant equal to 1. The lunar nearside consists primarily of maria and highlands, with dielectric constants of 3.42 and 4.13, respectively \cite{Gao:2023}. For the calculations presented here, an average value of $\epsilon=3.78$ is adopted. The emitted power per unit surface area from a lunar region with dielectric constant $\epsilon$ is given by \cite{Jaeckel:2013eha}
\begin{equation}
I_0=\chi^2\rho_\mathrm{DM}\sin(\alpha)^2\left({n-1\over n}\right)^2 ~~ , \label{eq:int_1}
\end{equation}
where $\chi$ is the photon-dark photon kinetic mixing parameter, $\rho_\mathrm{DM}$ is the dark matter energy density, assuming that all of the dark matter is composed of dark photons, $\alpha$ is the angle between the dark photon polarization and the vector normal to the surface, and $n=\sqrt{\epsilon}$ is the refractive index of the lunar material. The emitted radiation is concentrated in a spectral line centered at the frequency $\nu=m_{\gamma'}/(2\pi)$, where $m_{\gamma'}$ is the dark photon mass. The intrinsic spectral width is approximately $\delta\nu\sim10^{-6}\nu$.

For a randomly oriented dark photon polarization, the polarization-dependent factor in Eq. (\ref{eq:int_1}) can be averaged over all directions. If the dark photon field instead has a preferred polarization direction, the resulting signal can exhibit a modulation with a period of approximately 13.66 days, corresponding to half the lunar rotation period, provided that the dark photon field is not co-rotating with the solar system. The calculations below assume a randomly oriented dark photon field, for which
$\left<\sin(\alpha)^2\right>=2/3$.

For a perfectly flat interface, the emitted electromagnetic radiation is directed nearly perpendicular to the surface. The deviation from the surface-normal direction is of order the dark matter velocity, $v\sim10^{-3}$, while the corresponding angular width is set by the dark matter velocity dispersion, $\delta v\sim10^{-3}$ \cite{Jaeckel:2015kea,Jaeckel:2013sqa}. Surface roughness introduces a much broader angular distribution because different local surface slopes redirect the radiation in different directions. For the lunar surface, this roughness-induced angular spread dominates over the angular scales associated with the dark matter velocity. The effect of lunar surface roughness must consequently be incorporated into the angular emission distribution used to estimate the detectable signal.

\begin{figure}[t]
  \includegraphics[width=1\linewidth]{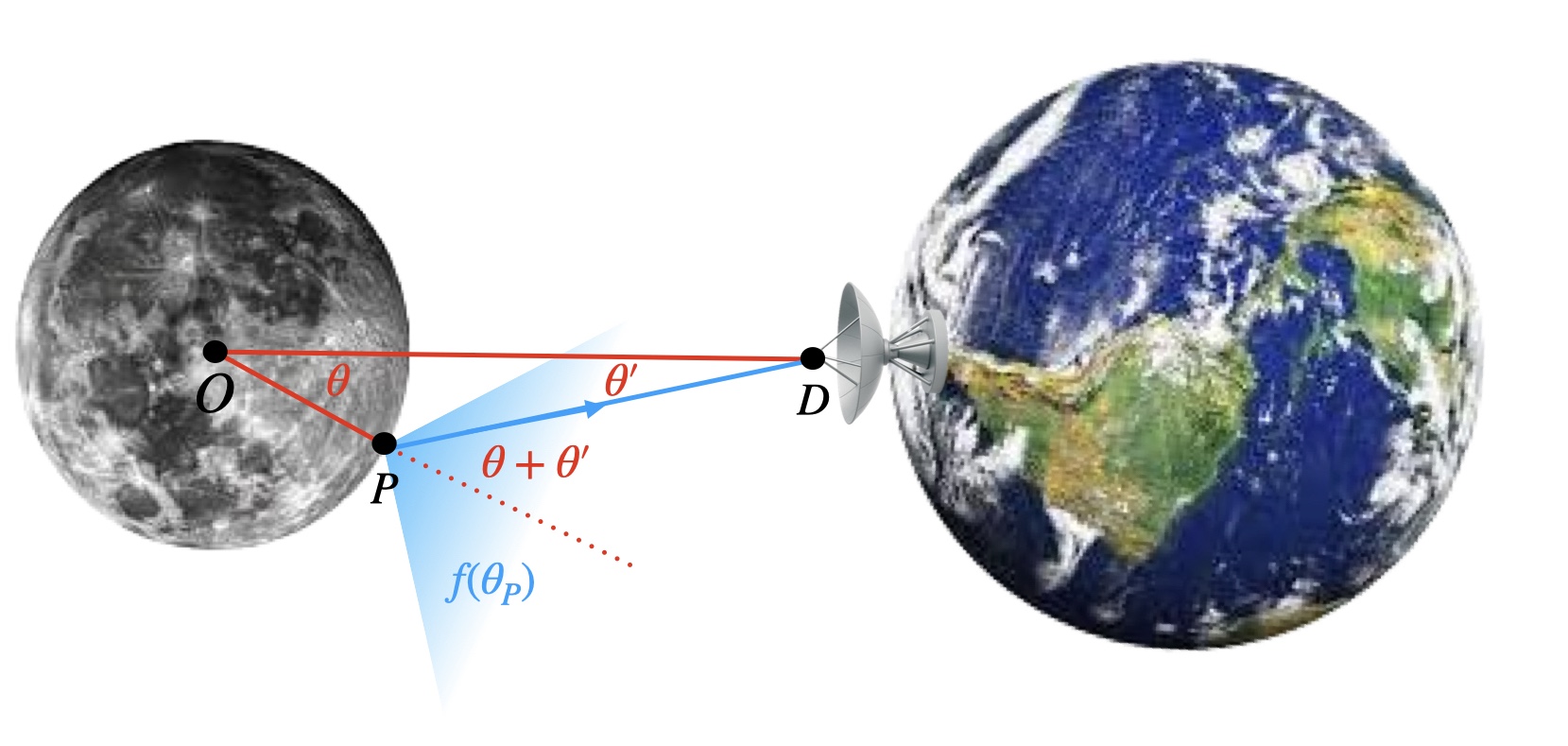}
\caption{A schematic of the proposed detection mechanism, where part of the radiation emitted from a point $P$ on the Moon is collected by an antenna located on Earth.} \label{fig:scheme}
\end{figure}

Figure~\ref{fig:scheme} illustrates the geometric setup. The point~$O$ denotes the center of the Moon, $D$ denotes the detector, and $P$ denotes an arbitrary point on the lunar surface from which radiation is emitted. The segment~$\overline{OP}$ corresponds to the lunar radius, $R_M = 1.74 \times 10^6\,\mathrm{m}$, while $\overline{DO}$ represents the distance between the detector and the lunar center, with an average value of $d = 3.78 \times 10^8\,\mathrm{m}$.

The position of $P$ is specified by the polar angle~$\theta$ and azimuthal angle~$\phi$ in a lunar-centered spherical coordinate system with origin at $O$. The polar axis is chosen along the segment~$\overline{OD}$. The distance between $P$ and the detector is then
$L = \sqrt{d^2 + R_M^2 - 2dR_M\cos\theta}$, as follows from the law of cosines. To describe the same geometry from the detector's perspective, an additional angle~$\theta'$ is defined between the segments~$\overline{DO}$ and~$\overline{DP}$. The two angular coordinates are related by $\sin\theta' = R_M\sin\theta / L$ and $\cos\theta' = (d - R_M\cos\theta) / L$.

The horizon condition determines the largest value of $\theta$ for which radiation emitted from $P$ can reach the detector along a direct line of sight. This limit corresponds to the lunar limb and is given by $\theta = \arccos(\gamma)$, or equivalently by $\theta' = \arcsin(\gamma)$ in the detector frame, where $\gamma = R_M/d = 4.53 \times 10^{-3}$.

Radiation emitted from each point $P$ is described by an angular distribution $f(\theta_P,\theta,\phi)$, where $\theta_P$ is the polar emission angle measured in a coordinate system centered at $P$, with the polar axis aligned with the segment~$\overline{OP}$. The angle $\theta_P$ ranges from $0$ to $\pi/2$ for radiation emitted into the outward hemisphere \footnote{For simplicity, the angular distribution is assumed to be azimuthally symmetric.}. The dependence of $f$ on $\theta$ and $\phi$ allows the angular emission distribution to vary across the lunar surface. For each surface position $(\theta,\phi)$, the distribution is normalized according to
$\int d\Omega_P f(\theta_P,\theta,\phi)=1$, where $d\Omega_P=\sin(\theta_P)d\theta_P d\phi_P$.

For radiation emitted from the surface element around $P$, only the component directed toward the detector contributes to the observed signal. In the chosen geometry, this direction corresponds to $\theta_P=\theta+\theta'$. The power received by the detector from the differential surface element is therefore
${\cal P}=I_0,dA\delta\Omega_P\left.f(\theta_P,\theta,\phi)\right|_{\theta_P=\theta+\theta'}$,
where $dA=R_M^2\sin(\theta)d\theta d\phi=R_M^2d\Omega$ is the differential lunar surface area and $\delta\Omega_P$ is the solid angle subtended by the antenna as viewed from $P$. The latter is determined by the projected area of the detector perpendicular to the line of sight and is given by $\delta\Omega_P=A'\cos(\theta')/L^2$, where $A'$ is the total area of the antenna. The corresponding flux density at the detector is therefore
\begin{equation}
dS_\nu={I_0R_M^2\over\Delta\nu L^2}\cos(\theta')d\Omega\left.f(\theta_P,\theta,\phi)\right|_{\theta_P=\theta+\theta'} ~~ , \label{eq:dS_1}
\end{equation}
where the effective signal bandwidth $\Delta\nu$ is set by the larger of the intrinsic signal bandwidth $\delta\nu$ and the receiver frequency resolution $\delta\nu_r$,
\begin{equation}
\Delta\nu=\max(\delta\nu,\delta\nu_r) ~~ . \label{eq:bandwidth_1}
\end{equation}

The total flux density is obtained by integrating Eq. (\ref{eq:dS_1}) over the portion of the lunar surface visible to the detector. If the contributing region extends to a polar angle $\theta_*$ in the lunar frame, the result can be expressed as
\begin{equation}
S_\nu={I_0\gamma^2\over\Delta\nu}J(\theta_*) ~~ , \label{eq:S_1}
\end{equation}
where
\begin{equation}
J(\theta_*)=\int_0^{\theta_*}d\theta\sin(\theta)\int_0^{2\pi}d\phi f(\theta,\phi) ~~ , \label{eq:J_1}
\end{equation}
with
$f(\theta,\phi)=\left.f(\theta_P,\theta,\phi)\right|_{\theta_P=\theta+\theta'}$.
In Eq. (\ref{eq:J_1}), corrections of order $\gamma$ and higher have been neglected. The quantity $J(\theta_*)$ therefore represents the fraction of the emitted power that is directed toward the detector after accounting for both the local angular emission pattern and the lunar viewing geometry. It constitutes the central quantity required to characterize the detectable signal and, consequently, to obtain realistic sensitivity projections.

\begin{figure}[t]
  \includegraphics[width=1\linewidth]{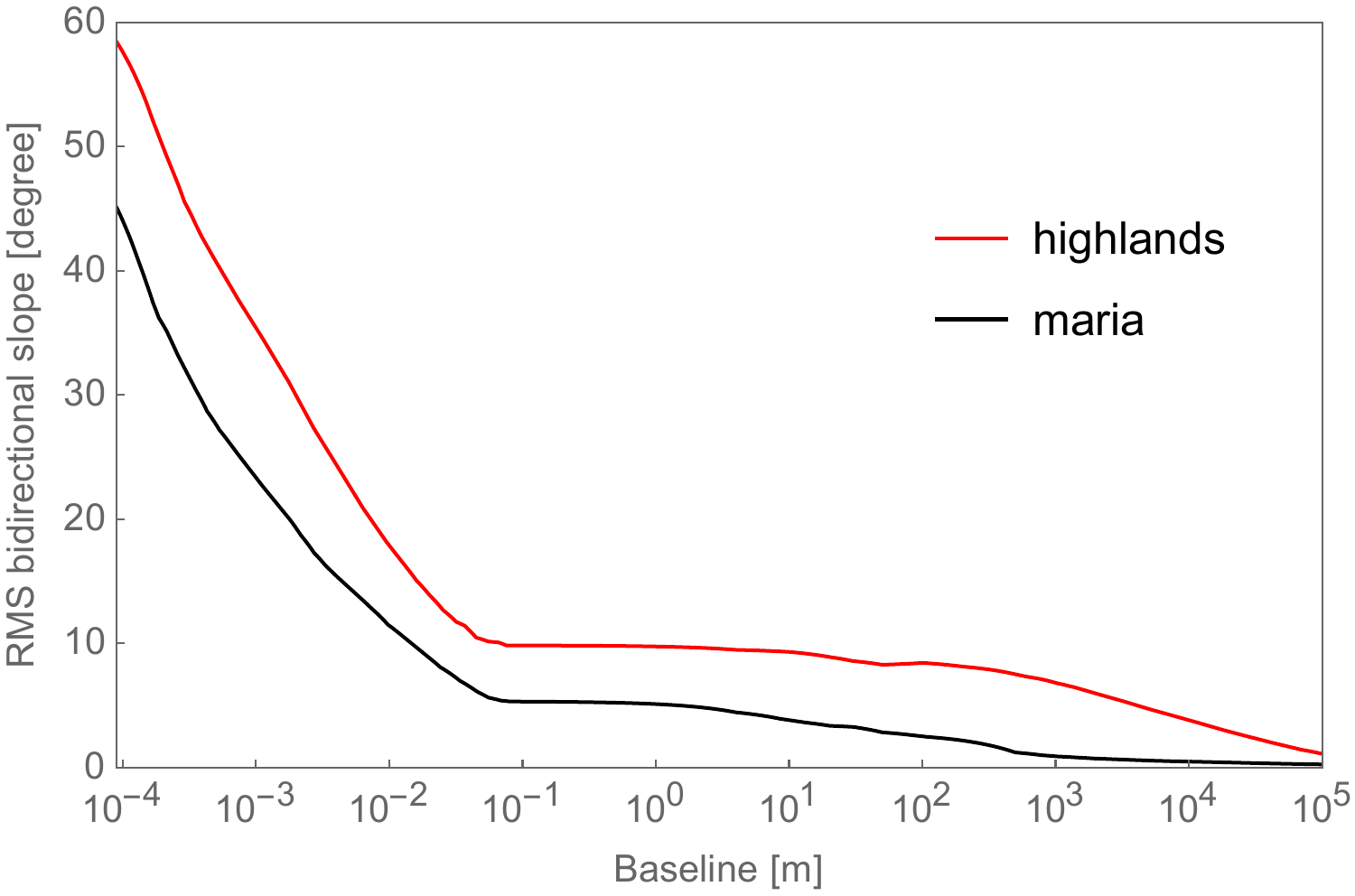}
\caption{RMS bidirectional slope for lunar maria (black) and highlands (red) at different baselines.} \label{fig:slope}
\end{figure}

{\it \textbf{Lunar surface roughness}.} To model radiation emitted from a rough terrain such as the lunar surface, a statistical description of the surface geometry must be incorporated into the angular distribution $f(\theta_P,\theta,\phi)$. A convenient characterization of surface roughness is provided by the RMS bidirectional slope at a given length scale, or baseline \cite{Shepard:2001,Rosenburg:2011}. For the Moon, measurements of the RMS bidirectional slope have been reported for different types of lunar terrain, including fresh, moderately degraded, and highly degraded craters, as well as regions characterized by large and small craters \cite{Guo:2021}. The lunar nearside, from which the signal considered here is emitted, is predominantly composed of maria and highlands. The corresponding RMS bidirectional slopes for these two terrain types, as a function of baseline length, are shown in Fig. \ref{fig:slope}. The values in the range $0.1-100\,\mathrm{mm}$ are taken from \cite{Helfenstein:1999slt}, while those in the range $4-10^5\,\mathrm{m}$ are extracted from \cite{Cai:2020mtr}. For the intermediate range, where measurements are unavailable, a simple extrapolation is adopted. This extrapolation is motivated by the approximately fractal character of lunar surface roughness reported in \cite{Turcotte:1987,Helfenstein:1999slt,Cao:2015,GlobalFractalMoon2026}.

\begin{table*}[t]
    \centering
    \begin{tabular}{|c|c|c|c|c|c|}
        \hline
                & FAST & GBT & SKA low & SKA mid & ALMA \\
        \hline\hline
       $\nu\,[\mathrm{GHz}]$ & 0.07 - 3 &  0.1 - 115 & 0.05 - 0.35 & 0.35 - 13.8 & 35 - 950 \\
       \hline
       $\delta\nu_r\,[\mathrm{kHz}]$ & 7.63 & 0.02 & 0.0141 & 0.21 & 31 \\
       \hline
       $N$ & 1  & 1 & 512 & 133 \& 64 & 54 \& 12 \\
       \hline
       $D\,[\mathrm{m}]$ & 500 & 100 & 35 & 15 \& 13.5 & 12 \& 7 \\
       \hline
       $\eta$ & 0.63  & 0.72 - 0.24 & 0.9 & 0.79 & 0.72 - 0.31 \\
       \hline
       $b_\mathrm{min}\,[\mathrm{m}]$ & - & - & 35 & 15 & 14.6 \\
        \hline
        $b_\mathrm{max}\,[\mathrm{km}]$ & - & - & 1 & 1 & 0.106 \\
        \hline
    \end{tabular}
    \caption{Telescope parameters considered in this work.}
    \label{tab:telescopes}
\end{table*}

For the angular distribution $f(\theta_P,\theta,\phi)$, the RMS bidirectional slope at a given baseline is identified with the RMS value of $\theta_P$, given by
\begin{equation}
\theta_P^\mathrm{RMS}=\sqrt{\left<\theta_P^2\right>}=\sqrt{\int d\Omega_P\,\theta_P^2\,f(\theta_P,\theta,\phi)} ~~ . \label{eq:thetaPRMS_1}
\end{equation}
For a dark photon of mass $m_{\gamma'}$, the relevant baseline is set by its wavelength, $\lambda=2\pi/m_{\gamma'}$. The RMS bidirectional slope relevant for a given dark photon mass is therefore associated with the RMS slope measured at a baseline equal to $\lambda$. More generally, the corresponding value is determined by the maximum RMS bidirectional slope at baselines larger than or equal to $\lambda$, since smaller-scale surface variations are not resolved by the wave\footnote{At length scales smaller than the wavelength, the wave does not resolve the corresponding surface slopes, and the radiation is emitted approximately in a single direction perpendicular to the surface averaged over that scale.}. Because the RMS bidirectional slope decreases monotonically with increasing baseline, this prescription is equivalent to evaluating the RMS bidirectional slope at a baseline equal to $\lambda$.

\begin{figure}[t]
  \includegraphics[width=1\linewidth]{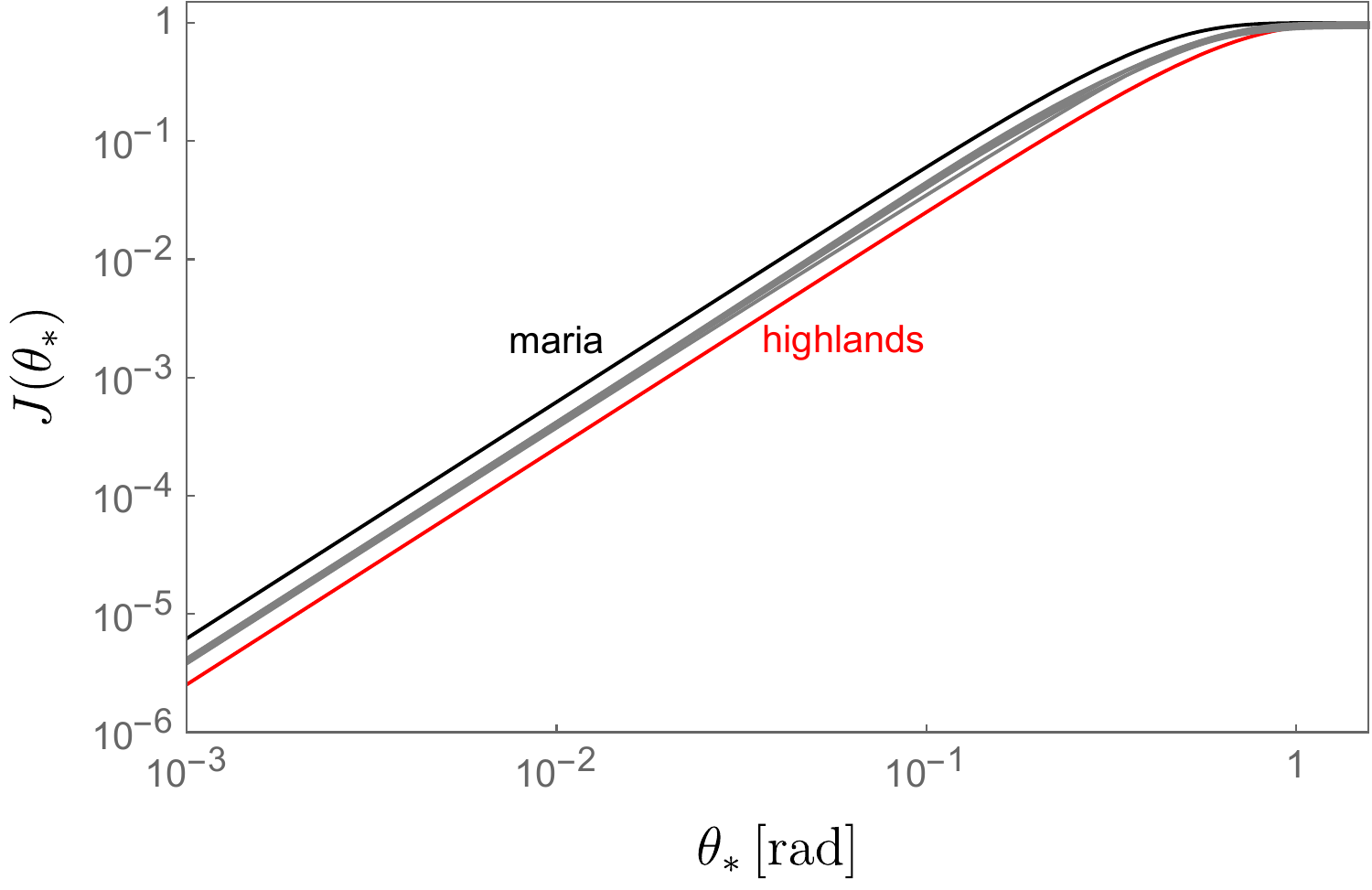}
\caption{$J(\theta_*)$ at 1 mm baseline assuming a homogeneous $\delta$ corresponding to lunar maria (black) and highlands (red). It also shows $J(\theta_*)$ evaluated with intermediate random functions for $\delta(\theta,\phi)$ (gray).} \label{fig:J}
\end{figure}

As shown in \cite{Helfenstein:1999slt}, the distribution of lunar surface slopes is approximately Gaussian. The angular distribution of the radio emission at each lunar coordinate $(\theta,\phi)$ is therefore modeled as
\begin{equation}
f(\theta_P,\theta,\phi)=C(\theta,\phi)e^{-{\theta_P^2\over2\delta(\theta,\phi)^2}} ~~, \label{eq:f_1}
\end{equation}
where $C(\theta,\phi)$ is the normalization constant,
\begin{equation}
C(\theta,\phi)=\left(\int d\Omega_P\,e^{-{\theta_P^2\over2\delta(\theta,\phi)^2}}\right)^{-1}.
\end{equation}

The width $\delta$ is related to the RMS bidirectional slope through Eq. (\ref{eq:thetaPRMS_1}). Specifically, the function $\theta_P^\mathrm{RMS}(\delta)$ is obtained from Eq. (\ref{eq:thetaPRMS_1}), and $\delta(\theta_P^\mathrm{RMS})$ is then determined by inverting this relation. The value of $\delta$ at each lunar coordinate is consequently fixed by the RMS bidirectional slope of the corresponding terrain at the relevant baseline.

For a given dark photon mass $m_{\gamma'}$, the RMS bidirectional slope at each point on the lunar nearside is expected to lie between the values associated with the two limiting terrain types shown in Fig. \ref{fig:slope}, namely, lunar maria and highlands. To assess the impact of spatial variations in the surface roughness, random realizations of $\delta(\theta,\phi)$ bounded by these two limiting values are considered and used to calculate $J(\theta_*)$. Figure \ref{fig:J} shows $J(\theta_*)$ for a baseline of $1\,\mathrm{mm}$, with a homogeneous $\delta$ that corresponds to maria (black line) and homogeneous $\delta$ that corresponds to highlands (red line). For comparison, the figure also shows $J(\theta_*)$ calculated with five random realizations of $\delta(\theta,\phi)$ (gray lines), with values bounded by the corresponding $1\,\mathrm{mm}$ baseline limits for maria and highlands. These realizations represent highly inhomogeneous lunar surfaces.

The resulting $J(\theta_*)$ for the inhomogeneous surfaces generally lies between the two homogeneous limiting cases, with deviations occurring only near the boundary $\theta_*\rightarrow\arccos(\gamma)$. For the sensitivity projections presented in this work, a homogeneous value of $\delta$ corresponding to the highlands case shown in Fig. \ref{fig:slope} is therefore adopted. This choice provides a conservative estimate by avoiding an overestimate of the expected signal. A more detailed spatially varying model for $\delta(\theta,\phi)$ could be incorporated in future work to account for the heterogeneous nature of the lunar surface and obtain a more accurate prediction of the signal.

{\it \textbf{Detection with radio telescopes}.} Sensitivity projections are considered for single-dish radio telescopes, such as the Five-hundred-meter Aperture Spherical Telescope (FAST) and the Green Bank Telescope (GBT), as well as antenna arrays, including the Atacama Large Millimeter/submillimeter Array (ALMA) and the future Square Kilometre Array (SKA). For the antenna arrays, both single-dish and interferometric observing modes are considered.

Several parameters are relevant for characterizing the response of a radio telescope, including the angular resolution, field of view (FoV), and effective area. For a single-dish telescope, as well as for an array of identical antennas operated in single-dish mode, the angular resolution and FoV are set by the diffraction limit and can be characterized by the same expression, although they have different physical meanings,
\begin{equation}
\theta_\mathrm{FoV}=\theta_\mathrm{res}\approx1.22{\lambda\over D} ~~ , \label{eq:FoV_1}
\end{equation}
where $\lambda$ is the wavelength and $D$ is the diameter of the dish. In interferometric mode, the angular resolution, also referred to as the synthesized beam, is determined by the longest baseline $b_\mathrm{max}$ and is approximately given by
$\theta_\mathrm{synth}\approx1.22{\lambda\over b_\mathrm{max}}$.
The FoV, on the other hand, is determined by the primary beam of the individual antennas and is therefore set by the dish diameter rather than by the longest interferometric baseline. For an array of identical dishes, it can be approximated as
$\theta_\mathrm{FoV}\approx1.22\lambda/D$.

For a single-dish telescope with physical collecting area $A_d$, the effective area is
\begin{equation}
A_\mathrm{eff}=\eta A_d ~~ , \label{eq:Aeff_1}
\end{equation}
where $\eta$ is the dish efficiency. For an array of $N$ identical dishes, the effective area entering the sensitivity estimate depends on the observing mode. In the single-dish mode and interferometric mode, respectively, it is given by
\begin{equation}
A_\mathrm{eff}=
\begin{cases}
\sqrt{Nn_p}\eta A_d ~~ , & \mathrm{single-dish}
\\
\sqrt{N(N-1)n_pn_\mathrm{pix}\over2}\eta A_d ~~ , & \mathrm{interferometer}
\end{cases} \label{eq:Aeff_2}
\end{equation}
where, in interferometric mode, $n_\mathrm{pix}$ denotes the number of independent synthesized beams contained within the primary beam. It can therefore be approximated by
$n_\mathrm{pix}=(\theta_\mathrm{FoV}/\theta_\mathrm{synth})^2$.
A summary of the relevant parameters for the radio telescopes considered here is given in Table \ref{tab:telescopes}. The corresponding information is taken from \cite{Nan:2011um,Jiang:2020fst} for FAST, \cite{GBTog,GBTpg} for GBT, \cite{SKAtel,Bonaldi:2024ska,Wayth:2023aeb} for SKA, and \cite{ALMAdocs} for ALMA.

\begin{figure}[t]
  \includegraphics[width=1\linewidth]{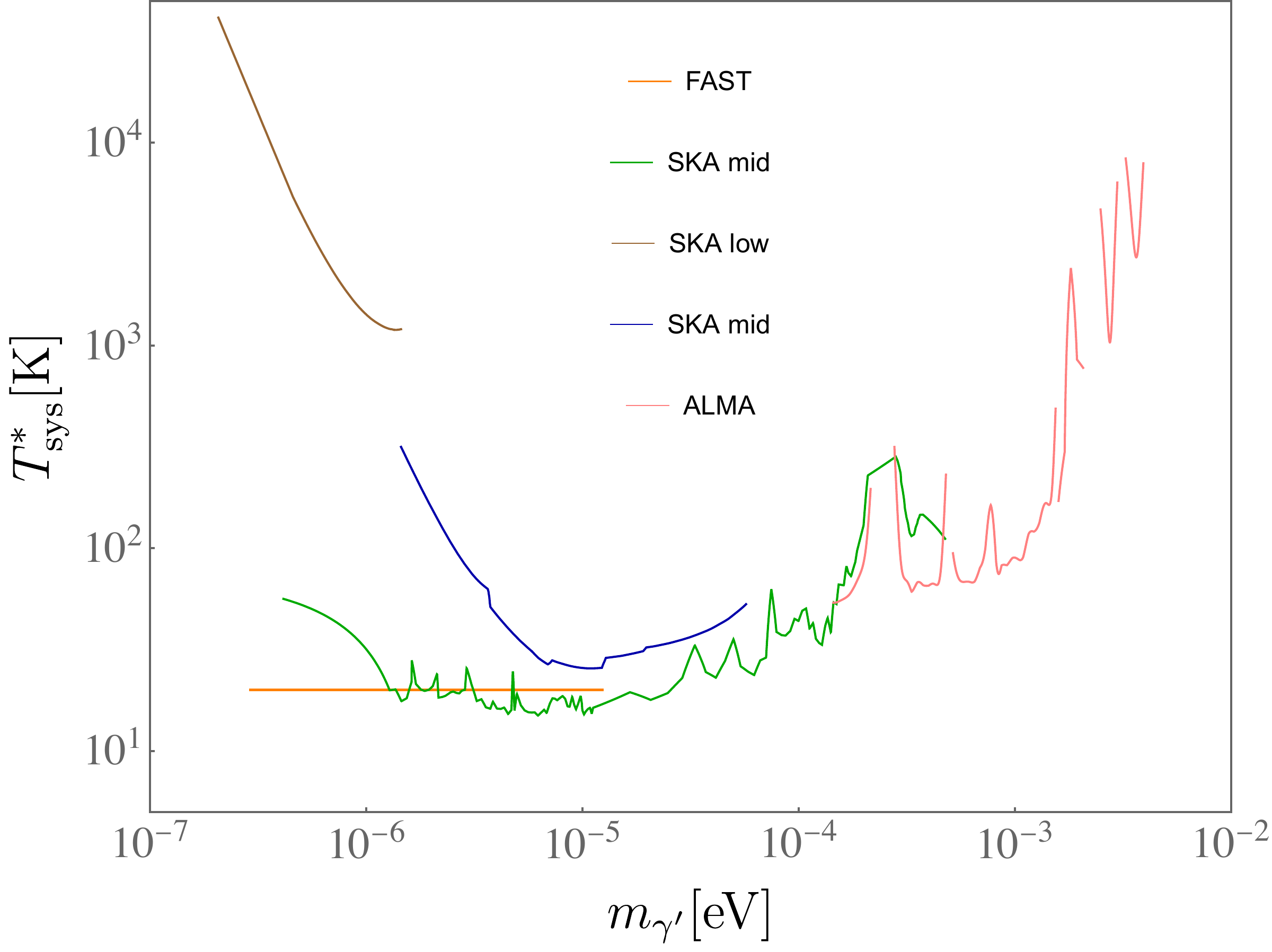}
\caption{System temperature $T_\mathrm{sys}^*$ for the radio telescopes used in this work.} \label{fig:Tsys}
\end{figure}

\begin{figure}[!t]
\includegraphics[width=1\linewidth]{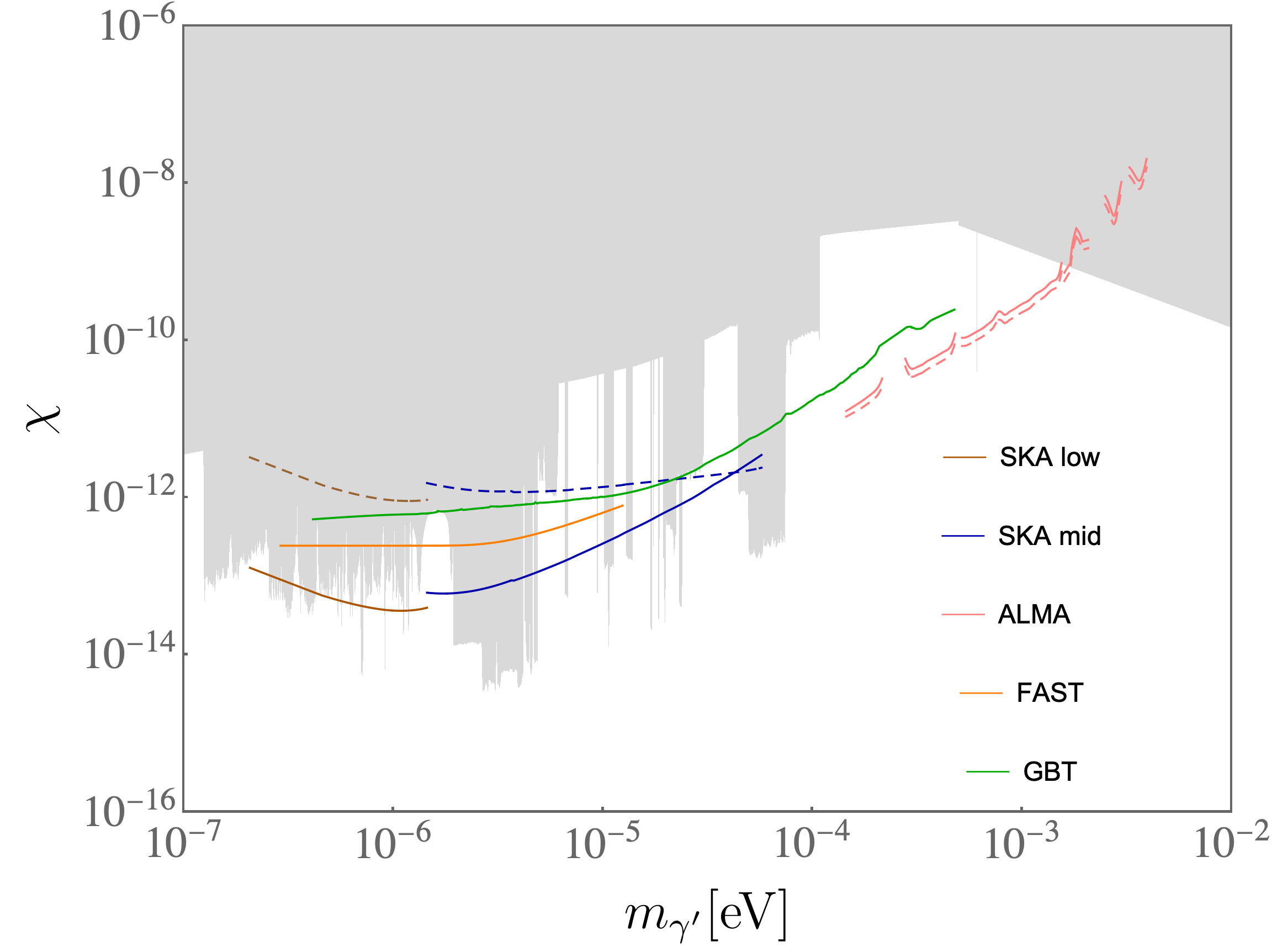}
\caption{Sensitivity projection for all the radio telescopes proposed in this work with a total observation time of 100 hours. For antenna arrays (SKA and ALMA), the dashed line corresponds to single-dish observation mode while the solid line to interferometer mode. The gray area corresponds to current constraints \cite{ohareaxionlimitsdp2020}.} \label{fig:sens}
\end{figure}

The minimum detectable flux density is determined by the radiometer equation,
\begin{equation}
S_\mathrm{min}={2T_\mathrm{sys}\over A_\mathrm{eff}\sqrt{n_p\Delta\nu t}} \label{eq:Smin}
\end{equation}
where $T_\mathrm{sys}$ is the system noise temperature, $A_\mathrm{eff}$ is the effective collecting area, $t$ is the total observation time, and $\Delta\nu$ is the effective signal bandwidth defined in Eq. (\ref{eq:bandwidth_1}). The parameter $n_p$ denotes the number of measured polarizations and is taken to be $n_p=2$. The sensitivity projections for the kinetic mixing parameter $\chi$ as a function of the dark photon mass $m_{\gamma'}$, shown in Fig. \ref{fig:sens}, are obtained by imposing the condition
\begin{equation}
S_\nu=S_\mathrm{min} ~~ . \label{eq:sens_1}
\end{equation}

The typical system temperature $T_\mathrm{sys}^*$ of each radio telescope considered here is shown in Fig. \ref{fig:Tsys}, where the data was taken from \cite{Nan:2011um} for FAST, \cite{GBTpg} for GBT, \cite{SKAtel} for SKA and the sensitivity calculator website for ALMA \cite{ALMAdocs2}. For observations of the Moon, an additional contribution from lunar thermal emission must be included in the total system temperature. At radio frequencies, the dominant component of lunar radiation is thermal emission produced by solar heating of the lunar surface. Because the Moon lacks a substantial atmosphere capable of retaining heat, its surface temperature varies significantly between lunar day and night. At the equator, where these variations are most pronounced, surface temperatures range from approximately $100$ to $400\,\mathrm{K}$ \cite{Paige2010Diviner,Zhang2022ChangE4,Williams2017Diviner}. Since the observations considered here are performed during lunar night, the contribution from lunar thermal emission is taken to be $T_M=100\,\mathrm{K}$. The total system temperature is therefore
\begin{equation}
T_\mathrm{sys}=T_\mathrm{sys}^*+T_M ~~ . \label{eq:Tsys}
\end{equation}

The total flux density $S_\nu$ is obtained by integrating Eq. (\ref{eq:dS_1}) over the portion of the lunar surface visible to the detector. The maximum angular extent of the lunar surface accessible to the antenna is characterized in the detector frame by
\begin{equation}
\theta_*'=\min(\theta_d,\arcsin(\gamma)) ~~ ,
\end{equation}
where $\theta_d=\theta_\mathrm{FoV}$ for a single-dish telescope and for an array operating in single-dish mode, while $\theta_d=\theta_\mathrm{synth}$ for an array operating in interferometric mode. The second condition accounts for the finite angular size of the Moon as viewed from the detector. In the lunar frame, the corresponding visible region is characterized by the angle $\theta_*$, which is obtained from $\theta_*'$ using the trigonometric relations described previously. As $\theta_*'$ and $\gamma$ are small, a very good approximation is
\begin{equation}
\sin(\theta_*)\approx{\theta_*'\over\gamma} ~~ . \label{eq:theta_star}
\end{equation}

{\it \textbf{Results and discussion}.} Sensitivity projections for an observation time of $t=100\,\mathrm{hr}$ are shown in Fig. \ref{fig:sens}. The projected sensitivities of SKA and FAST extend into previously unexplored regions of parameter space, including regions between existing constraints from haloscope searches. GBT and ALMA, on the other hand, provide sensitivity to an additional, relatively isolated region of unexplored parameter space at higher dark photon masses. At high masses, antenna arrays operated in single-dish mode provide greater sensitivity than in interferometric mode. This behavior results from the smaller synthesized beams in interferometric observations, which restrict the portion of the lunar surface contributing to the detected signal.

This work presents a detection strategy for dark photon dark matter based on the Moon acting as a dielectric converter. The electromagnetic flux density radiated from the lunar surface has been calculated while accounting for the effects of surface roughness over a broad range of length scales. Sensitivity projections have been obtained for several current and future radio telescopes, showing that this approach can probe regions of dark photon parameter space that are not accessible to existing searches.

The estimates presented here have been constructed conservatively to avoid overestimating the expected sensitivity. In particular, the angular distribution of the emitted radiation is modeled using the RMS bidirectional slope corresponding to the lunar highlands, rather than the smaller slopes associated with the maria. This choice results in a broader angular emission distribution and, consequently, lower projected sensitivity. A more detailed treatment could incorporate the spatially varying values of $\delta(\theta,\phi)$ across the lunar nearside, as well as the corresponding local variations in surface roughness. Such an analysis is left for future work and could provide a more accurate characterization of the expected signal and further refine the sensitivity projections presented here.

\bibliographystyle{JHEP}
\bibliography{references}

\end{document}